\documentclass{ifacconf}
\usepackage{xcolor}
\usepackage{subfigure}
\usepackage{graphicx} 
\usepackage{amsmath}
\usepackage{placeins}
\usepackage{amssymb}
\usepackage{natbib}        
\usepackage[nolist]{acronym}

\usepackage{cases}
\usepackage{mathtools}
\usepackage{stfloats}
\usepackage{comment}
\usepackage{textcomp}
\usepackage{multirow, booktabs}
\usepackage{lipsum}
\usepackage{cite}
\usepackage{subfigure}
\usepackage{float}
\usepackage{nomencl}
\usepackage{wasysym}
\usepackage{url}
\usepackage{empheq}

\usepackage{enumitem}
\usepackage[nolist]{acronym}
\usepackage{placeins} 
\newcommand{\circnum}[1]{\textcircled{\raisebox{-0.05pt}{\scriptsize #1}}}

\begin{acronym}
    \acro{DSO}{Distribution System Operator}
    \acro{RL}{Reinforcement Learning}
    \acro{PCNN}{Physically Consistent Neural Network}
    \acro{NN}{Neural Network}
    \acro{UMAR}{Urban Mining and Recycling}
    \acro{RB}{Rule-based}
    \acro{RES}{Renewable Energy Source}
    \acro{DSM}{Demand-Side Management}
    \acro{RBC}{Rule-based Control}
    \acro{MPC}{Model Predictive Control}
    \acro{DRL}{Deep Reinforcement Learning}
    \acro{AI}{Artificial Intelligence}
    \acro{ML}{Machine Learning}
    \acro{SRL}{Safe Reinforcement Learning}
    \acro{HVAC}{Heating, Ventilation, and Air Conditioning}
    \acro{DNN}{Deep Neural Network}
    \acro{DDPG}{Deep Deterministic Policy Gradient}
    \acro{BAU}{Business-as-usual}
    \acro{DQN}{Deep Q-Network}
    \acro{RASF}{Real-time Adaptive Safety Filter}
    \acro{DP}{Dynamic Programming}
    \acro{DER}{Distributed Energy Resource}

\end{acronym}  
\acresetall  
\begin{document}
\begin{frontmatter}

\title{Safe Deep Reinforcement Learning for Building Heating Control and Demand-side Flexibility} 


\author[First]{Colin Jüni} 
\author[First]{Mina Montazeri} 
\author[Second]{Yi Guo}
\author[First]{Federica Bellizio}
\author[Third]{Giovanni Sansavini}
\author[First]{Philipp Heer}

\address[First]{Urban Energy Systems Laboratory, Swiss Federal Laboratories for Materials
Science and Technology, Dubendorf, Switzerland.}
\address[Second]{School of Automation, Beijing Institute of Technology, Haidian District, Beijing, China}
\address[Third]{ Reliability and Risk Engineering Laboratory, ETH Zürich, CH-8092, Switzerland. }

\begin{abstract}
Buildings account for approximately 40\,\% of global energy consumption, and with the growing share of intermittent renewable energy sources, enabling demand-side flexibility, particularly in heating, ventilation and air conditioning systems, is essential for grid stability and energy efficiency. This paper presents a safe deep reinforcement learning-based control framework to optimize building space heating while enabling demand-side flexibility provision for power system operators. A deep deterministic policy gradient algorithm is used as the core deep reinforcement learning method, enabling the controller to learn an optimal heating strategy through interaction with the building thermal model while maintaining occupant comfort, minimizing energy cost, and providing flexibility. 
To address safety concerns with reinforcement learning, particularly regarding compliance with flexibility requests, we propose a real-time adaptive safety-filter to ensure that the system operates within predefined constraints during demand-side flexibility provision. 
The proposed real-time adaptive safety filter guarantees full compliance with flexibility requests from system operators and improves energy and cost efficiency --- achieving up to 50\,\% savings compared to a rule-based controller --- while outperforming a standalone deep reinforcement learning-based controller in energy and cost metrics, with only a slight increase in comfort temperature violations.
\end{abstract}
\end{frontmatter}
\section{Introduction}
\label{Sec:Introduction}
In 2022, the building sector accounted for 34\% of global energy demand and 37\% of the energy- and process-related carbon dioxide emissions \citep{buildingenergy2,buildingenergy1}, with consumption projected to rise by 50\% in the next 30 years according to the International Energy Agency \citep{2050building}. Meanwhile, the energy landscape is evolving, with a growing share of intermittent \acp{RES} and \acp{DER}, alongside the gradual phase-out of conventional power plants. These developments introduce operational uncertainty and pose critical challenges to power system operation \citep{ostergaard}. Simultaneously, advances in digitalization open new opportunities to leverage demand-side flexibility to support electrical system operation \citep{DSMinPowerSystems}, reducing reliance on conventional power generation, thus mitigating the need for costly grid reinforcements \citep{gridreinforcment}. In the building sector, \ac{DSM} programs are becoming increasingly popular due to the integration of energy management systems, which aim to optimize net energy consumption and enhance integration into a more flexible and low-carbon energy system \citep{DSMreview}.

To effectively control building energy systems, different control strategies are required to manage \ac{HVAC} systems in a dynamic environment.
Conventionally, \ac{RBC} relies on predefined rules and static thresholds to manage building energy systems. While widely adopted but simple, these controllers struggle to adapt to fluctuating energy prices, occupancy patterns, or renewable energy availability \citep{HVACcontrolreview}. To improve efficiency, \ac{MPC} has been introduced, leveraging forecasted data to optimize control decisions. However, its reliance on accurate system models and high computational complexity limits its real-time adaptability and scalability \citep{complexbuildings}. In contrast, \ac{RL} provides a data-driven alternative, autonomously learning optimal control policies by interacting with an environment. \ac{DRL} further enhances this capability by using deep neural networks to handle complex, high-dimensional state and action spaces, making it particularly suited for dynamic building energy management \citep{AplicationsRLBuilding}.

Numerous studies have leveraged \ac{DRL} for building energy management to improve thermal comfort and reduce energy use \citep{gupta,SILVESTRI2024123447,wei,kazmi}.
Gupta et al. \citep{gupta} implemented a \ac{DRL}-based heating controller that penalizes both comfort violations and energy costs, demonstrating up to 30\% improvement in comfort and 12\% reduction in energy costs compared to traditional thermostats. 
Similar results were reported by Silvestri et al. \citep{SILVESTRI2024123447}, Wei et al. \citep{wei}, and Kazmi et al. \citep{kazmi}, who demonstrated the effectiveness of \ac{DRL} in reducing energy consumption and temperature violations in \ac{HVAC} control scenarios, including water temperature control in residential buildings.\\
While \ac{DRL} has shown strong potential for managing complex, dynamic building environments \citep{surveybuildingRL}, its exploratory nature raises concerns about safety and constraint satisfaction, as it may lead to policy violations that impact occupant comfort or energy flexibility provision. This is particularly critical in \ac{DSM} programs, where buildings must respond to flexibility requests from a power system operator by imposed energy constraints  \citep{RLBuildingsoppandchallenges,zhao2023statewisesafereinforcementlearning}. 
In parallel, numerous control strategies beyond reinforcement learning have been explored to enable energy flexibility provision from buildings, such as load shifting techniques \citep{YANG2024100171}. Zhengguang et al. \citep{LIU2023100149} reviewed building energy flexibility and showed that the thermal inertia of a building plays a major role when considering different \ac{DSM} scenarios. Zhang et al. \citep{ZHANG2022100099} investigated an \ac{MPC}-based approach for flexibility provision, demonstrating the ability of \ac{MPC} to provide a multitude of flexibility services, including load shedding, shifting and load tracking. Friansa et al. \citep{10781388} investigated a \ac{DRL}-based controller to enhance energy flexibility in a building equipped with photovoltaic generation and a flexible \ac{HVAC} system, improving flexibility while maintaining occupant comfort However, many of these approaches either do not explicitly enforce energy constraints or depend on accurate system models, which may limit scalability across diverse building types.\\
To address the lack of safety guarantees, \ac{SRL} techniques have been developed and integrated with safety filters and constraint-handling mechanisms to ensure control policies remain within operational limits \citep{gu2024reviewsafereinforcementlearning}. These approaches have been successfully applied in domains such as robotics, autonomous drones, and video games, where safety violations can result in problematic failures \citep{coptersSRL,INAMDAR2024100810}.
By incorporating explicit energy constraints, \ac{SRL} enables buildings to dynamically adjust their heating strategies to fulfill flexibility requests without violating contractual obligations with power system operators. These safety-aware \ac{DRL} frameworks allow buildings to act as active participants in balancing and congestion mechanisms, supporting a more resilient and decentralized energy system. In contrast to model-based control methods - which require accurate system models \citep{coptersSRL,SRL_videogame} - model-free \ac{SRL} offers a promising alternative, enabling the controller to learn and adapt strategies while explicitly ensuring compliance energy constraints.\\
Despite its strong potential, the application of \ac{SRL} in building energy management remains largely unexplored, particularly in the context of real-time flexibility provision. Huo et al. \citep{huo2024optimalmanagementgridinteractiveefficient} proposed a physics-inspired safe \ac{DRL} method for temperature control, using prior knowledge and hard steady-state rules to guarantee compliance with grid-interaction constraints. Paesschesoone et al. \citep{PAESSCHESOONE2024123507} proposed a safe \ac{DRL} controller with a predictive safety filter based on \ac{MPC} to enhance the efficiency and safety of existing energy flexibility controllers. Using a data-driven \ac{MPC} safety filter to adjust the \ac{DRL} agent's action, adherence of energy constraints can be guaranteed. Similarly, Wang et al. \citep{SRLbuilding} applied an \ac{MPC}-based filter layer to correct potentially unsafe states using an approximate building model. While these safe \ac{DRL} approaches show promising results, they rely on system identification and physical modeling, or handcrafted rules --- all of which limit their scalability and applicability across diverse building types.  

In this paper, we design and implement a safe \ac{DRL}-based controller for flexibility provision within an energy hub. To ensure safe operation without relying on physical models or prior system knowledge, we introduce a real-time adaptive safety filter that adjusts the \ac{DRL} agent’s actions based on real-time observations of room temperature and dynamic electricity price. For simulation, we model the thermal dynamics of the building using \acp{PCNN}, which integrates physical system knowledge with data-driven learning. 
The key contributions of this work are threefold:
\begin{enumerate} 
    \item A model-free real-time adaptive safety filter that requires no prior system knowledge or model identification. The filter adjusts control actions in real-time based on room temperature and electricity price to ensure compliance with flexibility requests from power system operators. 
    \item The integration of the safety filter into a \ac{DRL}-based control framework for demand-side flexibility in building space heating, enabling safe operation in real-time without compromising control performance. 
    \item A close-to-reality simulation using data from a living laboratory building\footnote{UMAR apartment unit at the NEST building on the Empa campus in Dübendorf, Switzerland.}, demonstrating the controller’s ability to optimize energy usage, ensure thermal comfort, and deliver flexibility services. 
\end{enumerate}
\section{Problem Formulation}
\label{Sec:Problem Formulation}
In this section, the control strategy for the flexible operation of a heat pump for space heating is presented. The controller minimizes the electricity costs while ensuring the constraints regarding system dynamics, thermal comfort, and flexibility provision.
\subsection{Optimization-Based Control}
\label{subsec:optbasedcontrol}
\label{Sec:Optimization Problem}
The control problem is formulated to determine optimal heat pump input \(u_t \in \mathbb{R}\) over the time horizon  \(t \in \mathcal{T}\), subject to system dynamics and operational constraints:

\begin{subequations} \label{eq:RSU_opt}
\begin{align}
    \min_{u_t} \quad & \mathcal{J}(u_t, \zeta(t)) \quad \label{eq:RSU_opt_a}, \tag{1a} \\
    \text{subject to} \quad & T_{t}^{\text{min}} \leq T_{\textrm{room},t} \leq T_t^{\textrm{max}}, \label{eq:RSU_opt_b} \tag{1b} \\
    & T_{\text{room}, t+1} = h(T_{\text{room},t}, u_t), \label{eq:RSU_opt_c} \tag{1c} \\
    & \sum_{t \in \mathcal{H}} f(u_t) \leq V(F), \label{eq:RSU_opt_d} \tag{1d}
\end{align}
\end{subequations}
\noindent where \(\mathcal{J}\) is the objective function, i.e. the electricity costs to be minimized, and \(\zeta(t)\) is the real-time electricity price at time \(t\). We use $T_{\text{room},t}$ to denote the indoor room temperature at timestep $t$, and \(T_t^{\text{min}}\) and \(T_t^{\text{max}}\) are the minimum and maximum temperature limits for thermal comfort. The system dynamics \(h(T_{\text{room},t}, u_t)\) capture the thermal behavior of the room, linking the current temperature and control input to the predicted room temperature at the next time step. This function depends on the building thermal model of the room. The term \(f(u_t)\) represents the energy consumption associated with the control input, which is accumulated over the duration of a flexibility provision time window $t\in \mathcal{H} \subseteq \mathcal{T}$. The constraint in \eqref{eq:RSU_opt_d} ensures that the total energy consumed within the flexibility time window does not exceed the allowable energy budget \(V(F)\), as defined by the flexibility provision message \(F\). This constraint introduces a global coupling across time steps, making the control problem temporally dependent for allocation of energy throughout the flexibility time window. More details and an explanation of flexibility constraints will be presented in Section~\ref{Sec:Reinforcement Learning}. 

\subsection{Control Architecture}
\label{Sec:Controller Design}
\begin{figure}[ht!]
\centering
\includegraphics[width=0.9\linewidth]{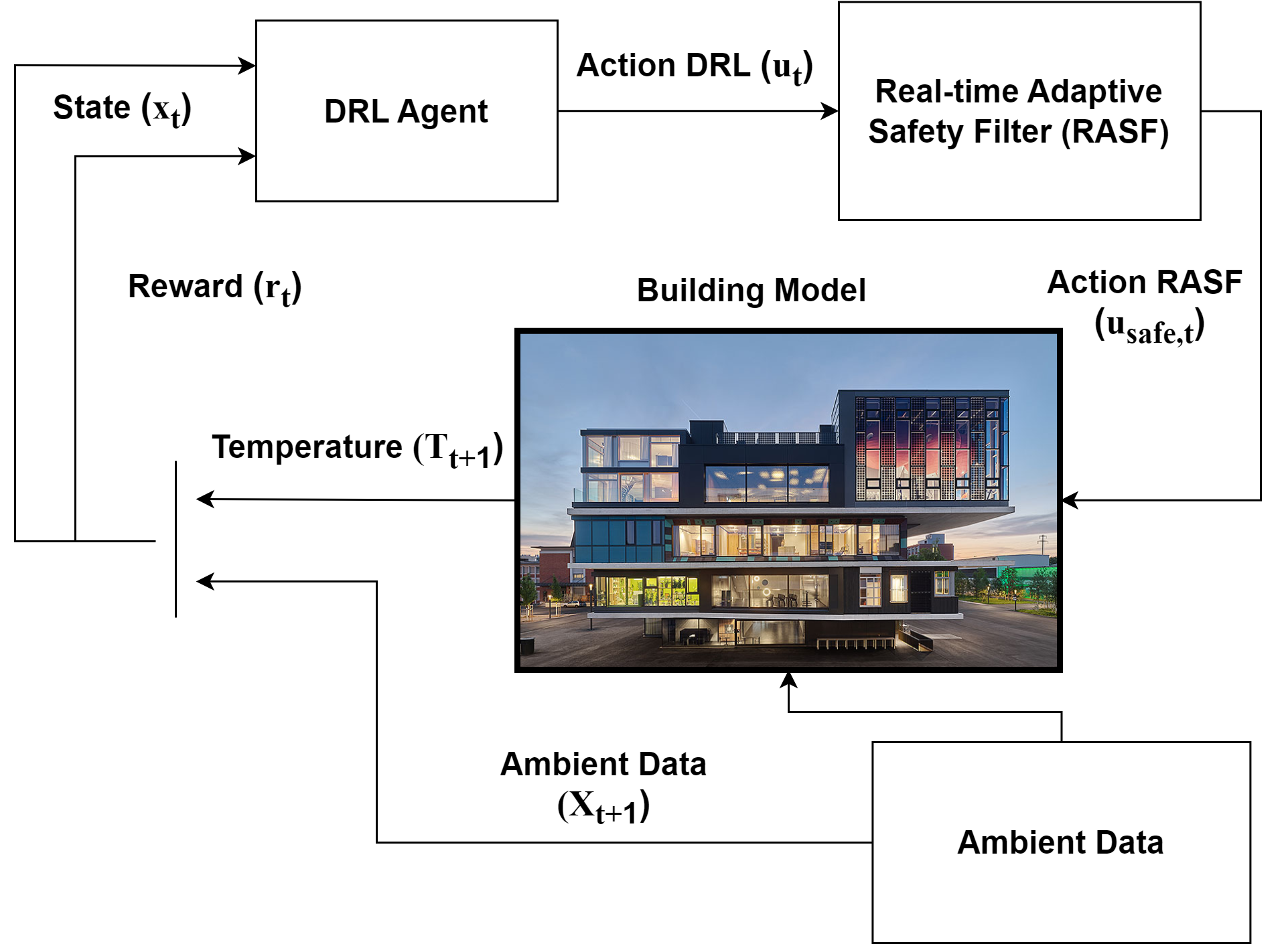}
\caption{Overview of the proposed safe DRL-based control scheme for space heating and demand-side flexibility. }
\label{Fig:SRLframework}
\end{figure}

The controller optimizes heat pump operation to ensure energy efficiency under varying ambient and operational conditions. 
To this end, we propose a model-free \ac{DRL} control architecture combined with a safety filter to ensure compliance with energy flexibility constraints. The overall design is shown in Fig. \ref{Fig:SRLframework}. 

The \ac{DRL} controller is the primary decision-making instance, computing optimal control action \(u_t\) based on the current state of the system \(x\). By interacting with the environment, the \ac{DRL} controller learns an optimal policy, maximizing the reward function to take an action \(u_t\) based on the state \(x\) and reward \(r\) at the previous time step.
The agent interacts with the building model, which uses a hybrid thermal model of a room to simulate the temperature at the next time step \(t+1\) based on the previous temperature and ambient data at timestep \(t\). Detailed mathematical description of the framework is provided in Section \ref{Sec:Reinforcement Learning}.

Although we consider the flexibility provision in the state and in the reward function of the \ac{DRL} controller, there is no guarantee that the energy flexibility requests are fulfilled at all times. Therefore, an additional safety measure is needed to fulfill the flexibility constraints at all times and thus ensure efficient operation of the grid.
The \ac{RASF}, introduced in Section \ref{Sec:Safety Filter}, is the critical component that guarantees all actions applied to the environment comply with the predefined flexibility constraints outlined in the flexibility provision message by adjusting the proposed \ac{DRL} action \(u_{t}\) to a safe control input \(u_{\text{safe},t}\).

\subsection{Thermal Modeling of Room Temperature}
\label{Sec:Thermal modeling of room temperature}

To simulate the thermal dynamics of a room, we use a hybrid modeling approach based on \acp{PCNN} \citep{PCNN1}. \acp{PCNN} combine a linear model incorporating physical knowledge and a non-linear data-driven neural network model to capture unforced dynamics that physical equations cannot quickly characterize. This structure enables accurate, interpretable temperature prediction while maintaining flexibility accross different buildings \citep{PiNN}.

The room temperature at the next timestep is modeled as $T_{\text{room},t+1} = D_{t+1} + P_{t+1}$, where $P_{t+1}$ captures the physical principles and $D_{t+1}$ accounts for the unforced dynamics learned from historical data via a \ac{NN}.
For more details of this hybrid model using a \ac{PCNN}, please refer to Di Natale et al. \citep{PCNN1}.

\subsection{Flexibility Provision}
\label{Sec:Flexibility provisions}
Flexible building operation refers to a building's ability to adjust its energy consumption patterns in response to signals or requests from the system operators \citep{ostergaard}. In our study, the controller incorporates flexibility provision messages by the operators while maintaining thermal comfort and minimizing electricity costs. 

\begin{figure}[b!]
    \centering
    \includegraphics[width=0.9\linewidth]{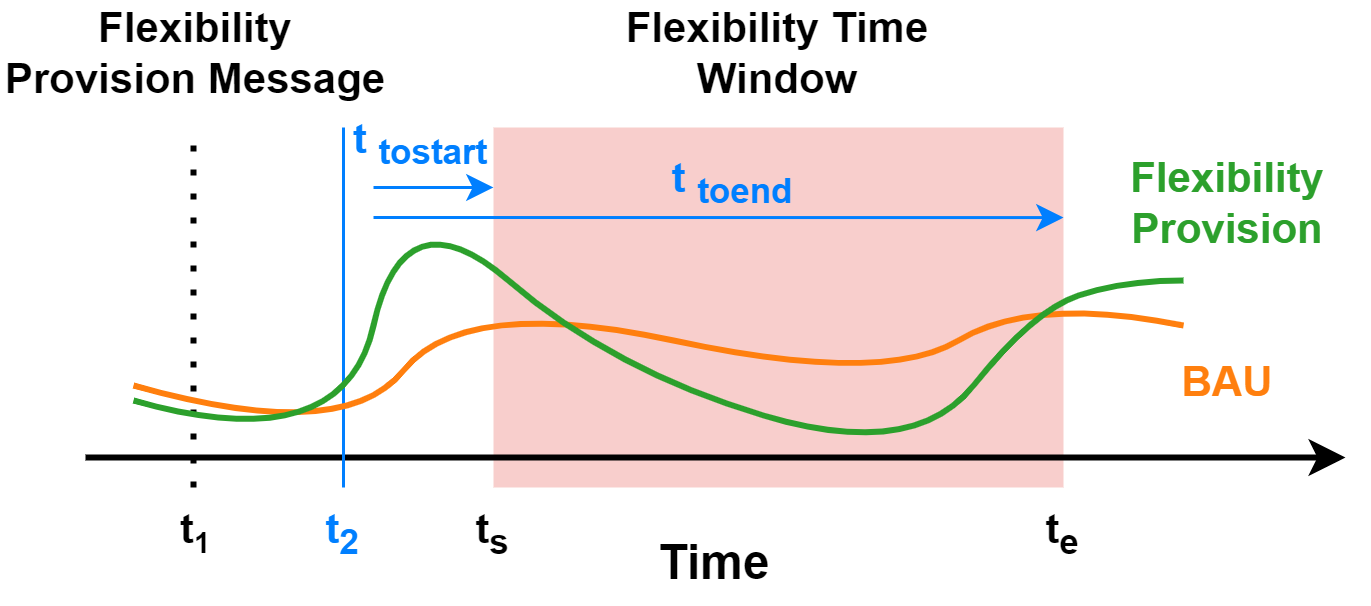}
    \caption{Timeline representation of a flexibility provision message and the associated flexibility time window with starting and ending times \(t_s\) and \(t_e\). The flexibility provision message is received at \(t_1\), with the states \(t_{\text{tostart}}\) and \(t_{\text{toend}}\) at time \(t_2\). The orange line shows the \ac{BAU} energy consumption trajectory, while the green line represents the flexibility provision trajectory.}
    \label{fig:flextimewindow}
\end{figure}

We consider flexibility provision messages in the form of \(F = (t_\text{s}, t_\text{e}, \phi_{\text{flexibility}})\), where \(t_\text{s}\) and \(t_\text{e}\) represent the starting and ending times of a flexibility provision time window $\mathcal{H}$, respectively, and \(\phi_{\text{flexibility}}\) denotes the flexibility factor. The flexibility factor \(\phi_{\text{flexibility}} \in [\phi_{\text{low}}, \phi_{\text{high}}]\)  specifies the energy consumption during the flexibility time window compared to the \ac{BAU} scenario without external intervention, as visualized in Fig. \ref{fig:flextimewindow}.

The deep reinforcement learning-based controller incorporates the flexibility provision messages \(F\) as part of the state space. This approach allows the \ac{DRL} agent to learn policies that naturally comply with flexibility constraints during training, such as preheating the room before a flexibility provision time window starts. However, this does not guarantee that flexibility constraints are fully satisfied in all scenarios. Thus, an additional safety filter is required to ensure strict compliance with flexibility constraints.

\section{Deep Reinforcement Learning Controller}
\label{Sec:Reinforcement Learning}

As mentioned above, the \ac{DRL} algorithm aims to determine the optimal control set points for the heat pump, specifically the valve opening level. 
Since we consider the valve opening as the \ac{DRL}'s action (\(u = [u_{t}]\)) and it is a continuous variable, we apply \ac{DRL} using a \ac{DDPG} algorithm which is well-suited for continuous action spaces. 
The \ac{DDPG} framework consists of two neural networks: a critic network, which estimates the value function (Q-value) for state-action pairs, and an actor network, which learns a deterministic policy to select actions given a state. We define the state of the \ac{DRL} agent at time $t$ as follows:
\begin{align}
    S_t = [ & I_{\text{solar},t}, T_{\text{amb},t}, T_{\text{room},t}, T_{\text{neigh},t}, X_{\text{time},t}  \notag \\
                   & u_{t}, c_{\text{heat},t}, t_{\text{tostart},t}, t_{\text{toend},t}, E_{\text{BAU},t}, E_t],
    \label{eq:state2}
\end{align}
where \(I_{\text{solar},t}\) represents the solar irradiation,
\(T_{\text{amb},t}\) is the ambient temperature, \(T_{\text{room},t}\) is the controlled room temperature, and \(T_{\text{neigh},t}\) is the temperature of the neighboring room, influencing heat transfer. 
The variable \(X_{\text{time},t}\) specifies the time of day, weekday, and season, while \(c_{\text{heat},t}\) denotes the heating or cooling scenario.
The terms \(t_{\text{tostart},t}\) and \(t_{\text{toend},t}\) define the flexibility provision window.
The \ac{BAU} energy \(E_{\text{BAU},t}\) is the amount of energy used without flexibility provision. The simulator computes the \ac{BAU} energy consumption by running the \ac{DRL} agent without considering flexibility constraints on historical data. 
The energy that has already been used during the flexibility time window is described by \(E_t\), so that the agent knows how much more energy it can use minimally or maximally.

The reward function helps the agent to maintain a balance between energy consumption and thermal comfort while fulfilling flexibility provisions. Thus, the reward function is given by:
\begin{align}
\label{eq:rewardfunction}
R_t = \beta \cdot R_{\text{temp},t} - \delta  \cdot R_{\text{price},t} - R_{\text{flex},t},
\end{align}
where $R_{\text{temp},t}$ penalizes deviations from predefined comfort temperature bounds, $ R_{\text{price},t}$ accounts for electricity costs, and $ R_{\text{flex},t}$ penalizes non-compliance with flexibility constraints. The coefficients \(\beta\) and \(\delta\) assign different weights to comfort violation and electricity cost. Each reward component is defined as:
\begin{align}
&R_{\text{temp},t} =  \max(T_{\text{room},t} - T^{\text{max}}_t, 0) - \min(T_{\text{room},t} - T^{\text{min}}_t, 0), \notag\\
&R_{\text{price},t} =   u_{t} \cdot \zeta(t), \notag\\
&R_{\text{flex},t} =  \alpha.\notag
\end{align}
The penalty term \(\alpha\) for flexibility constraints is defined as:
\begin{equation}
    \alpha =
    \begin{cases} 
        \alpha_{1}, & \text{if } E_t > E_{\text{BAU},t}, \\[6pt]
        \alpha_{2}, & \text{if } E_t > E_{\text{BAU},t} \text{ and } t_{\text{toend},t} = 0, \\[6pt]
        0, & \text{otherwise}.
    \end{cases}
\end{equation}
To train the neural networks in the \ac{DDPG} algorithm, we apply an actor-critic learning framework. The actor network, denoted as $\mu(S_t|\theta^\mu)$, represents the policy and determines the optimal valve opening level $u_t$ based on the given state $S_t$.  The parameter set $\theta^\mu$ consists of the weights and biases of the neural network that parametrizes the policy function $\mu$. The critic network, denoted as $Q(S_t,u_t|\theta^Q)$ assesses the long-term impact of selecting a specific valve opening level by estimating the expected future reward for a given state-action pair. The parameter set $\theta^Q$ contains the weights and biases of the critic network that approximates the Q-function. The Q-function represents the expected cumulative reward that the agent can achieve starting from a given state and taking a specific action, while following the current policy.
The actor network is trained to maximize the expected cumulative reward, ensuring that the valve operates efficiently while maintaining thermal comfort and meeting flexibility constraints. This is formulated as the following objective function:
\begin{equation} J(\theta^\mu) = \mathbb{E} \left[ \sum_{t=0}^{\infty} \gamma^t R_t \mid S_0, \mu \right], \end{equation}
where $\gamma$ is the discount factor that determines the importance of future rewards. To optimize this objective, the actor network is updated using the deterministic policy gradient:

\begin{equation} 
\nabla_{\theta^\mu} J \approx \mathbb{E} \left[ \nabla_u Q(S, u \mid \theta^Q) \Big|{u = \mu(S)} \nabla_{\theta^\mu} \mu(S \mid \theta^\mu) \right]. \end{equation}
The critic network is trained using the Bellman equation, which ensures the Q-value estimates align with the expected future rewards. The target Q-value is computed as:
\begin{equation}
y_t = R_t + \gamma Q(S_{t+1}, \mu(S_{t+1} \mid \theta^\mu) \mid \theta^Q). 
\end{equation}
The critic network minimizes the loss function:
\begin{equation}
L(\theta^Q) = \mathbb{E} \left[ (y_t - Q(S_t, u_t \mid \theta^Q))^2 \right]. 
\end{equation}
By continuously updating both networks, the \ac{DRL} controller learns to adjust the valve opening dynamically in response to ambient conditions, electricity prices, and flexibility constraints. This enables efficient heat pump operation while maintaining comfort and optimizing energy consumption.

\section{Real-time adaptive Safety Filter Design}
\label{Sec:Safety Filter}

The \ac{RASF} ensures that the actions proposed by the \ac{DRL} agent always fulfill the flexibility constraints. At each time, the agent proposes an action \(u_t\), which represents the opening factor of the valve of the heat pump. The proposed safety filter evaluates this action and adjusts it in a real-time manner, based on real-time conditions, to produce a safe control input \(u_{\text{safe},t}\), thereby guaranteeing compliance with the predefined flexibility constraints. The framework with the embedded \ac{RASF} is displayed in Fig. \ref{Fig:SRLframework}.

The proposed safety filter operates as follows: The remaining average safe action \(u_{\text{1},t}\) is computed based on the remaining energy budget \(E_{\text{remaining},t} = E_{\text{BAU},t} - E_t\) to fulfill the flexibility request and the remaining time \(t_{\text{toend},t}\) of the flexibility window, so that
\begin{equation}
    u_{\text{1},t} = \frac{E_{\text{remaining},t}}{t_{\text{toend},t}}.
    \label{eq:}
\end{equation}
Thus, the control input \(u_{\text{safe},t}\) precisely meets the flexibility request, as it is derived as the average input needed per timestep over the remaining duration of the flexibility window. 

To adapt to varying operational conditions, the safety filter adjusts the tolerance \(\tau_t\) in real-time, increasing it when greater flexibility is both necessary and beneficial. In particular, higher tolerance is required when the room temperature approaches its comfort limits, as the system needs more freedom to act and avoid violating thermal comfort constraints. At the same time, low electricity prices offer an opportunity to consume energy more freely without incurring high costs. By increasing \(\tau_t\) in these situations, the filter allows the controller to respond more aggressively when either comfort risk or economic benefit is high. This behavior is modeled through a weighted combination of two normalized indicators: thermal flexibility and price favorability. The tolerance is defined as:
\begin{equation}
    \tau_t = \tau_{\text{base},t} \cdot (w_1 \cdot\mu_{\text{temp},t} + (1 - w_1) \cdot\mu_{\text{price},t})^\text{1.5},
    \label{eq:tau}
\end{equation}
where \(\mu_{\text{temp},t} = \left( 1 - T_{\text{room},t} \right) \), 
\( \mu_{\text{price},t} = ( 1 - \zeta(t)) \).
The base tolerance \(\tau_{\text{base},t}\) is defined based on the position within the flexibility time window, decreasing as the flexibility period progresses. In the early stages, i.e., time steps near the start of the flexibility window \(t_s\) as illustrated in Fig.~\ref{fig:flextimewindow}, a higher tolerance is applied to allow more deviation and provide greater control flexibility. Toward the end of the window, the tolerance is gradually tightened to ensure full adherence to the cumulative flexibility constraint. The coefficient \(\mu_{\text{temp},t}\) is a temperature-based weight and \(\mu_{\text{price},t}\) is a cost-based weight. 
The indoor room temperature $T_{\text{room},t}$ and the electricity price $\zeta(t)$ are normalized between [0.1, 0.9] using min-max normalization and \(w_1\) is a weighting factor that balances the influence of the electricity price and the temperature, respectively.
This dynamic adaptation prevents violations while allowing the \ac{DRL} controller to explore efficient energy management strategies.


With the introduced tolerance \(\tau_t\), the maximal and minimal actions of the valve of the heat pump are defined by: 
\begin{align}
   & u_{\text{maximal},t} = \min \left( u_{\text{1},t} \cdot (1 + \tau_t), 1 \right),
   \notag \\
    &u_{\text{minimal},t} = \max \left( u_{\text{1},t} \cdot (1 - \tau_t), 0 \right).
    \label{eq:min}
\end{align}
The maximal and minimal control inputs defined by the \ac{RASF} account for scenarios where energy consumption needs to decrease or increase, respectively.

Finally, depending on the action proposed by the \ac{DRL} agent \(u_{t}\), the safety filter adjusts the action to \(u_{\text{safe},t}\), which is then fed to the building environment to ensure the flexibility request is fulfilled and the predefined energy flexibility is provided. For the case where energy consumption needs to decrease during the flexibility time window, \(\phi_{\text{flexibility}} \leq 1\), the action is defined by $ u_{\text{safe},t} = \min \left( u_{t}, u_{\text{maximal},t} \right)$,
Conversely, when energy consumption needs to increase,  \(\phi_{\text{flexibility}} \geq 1\), the action is given by $u_{\text{safe},t} = \max \left( u_{t}, u_{\text{minimal},t} \right)$.


\section{Case study}
\label{Sec:Case study}

\begin{figure}[t]
    \centering
    \subfigure[]{ 
        \includegraphics[width=0.225\textwidth]{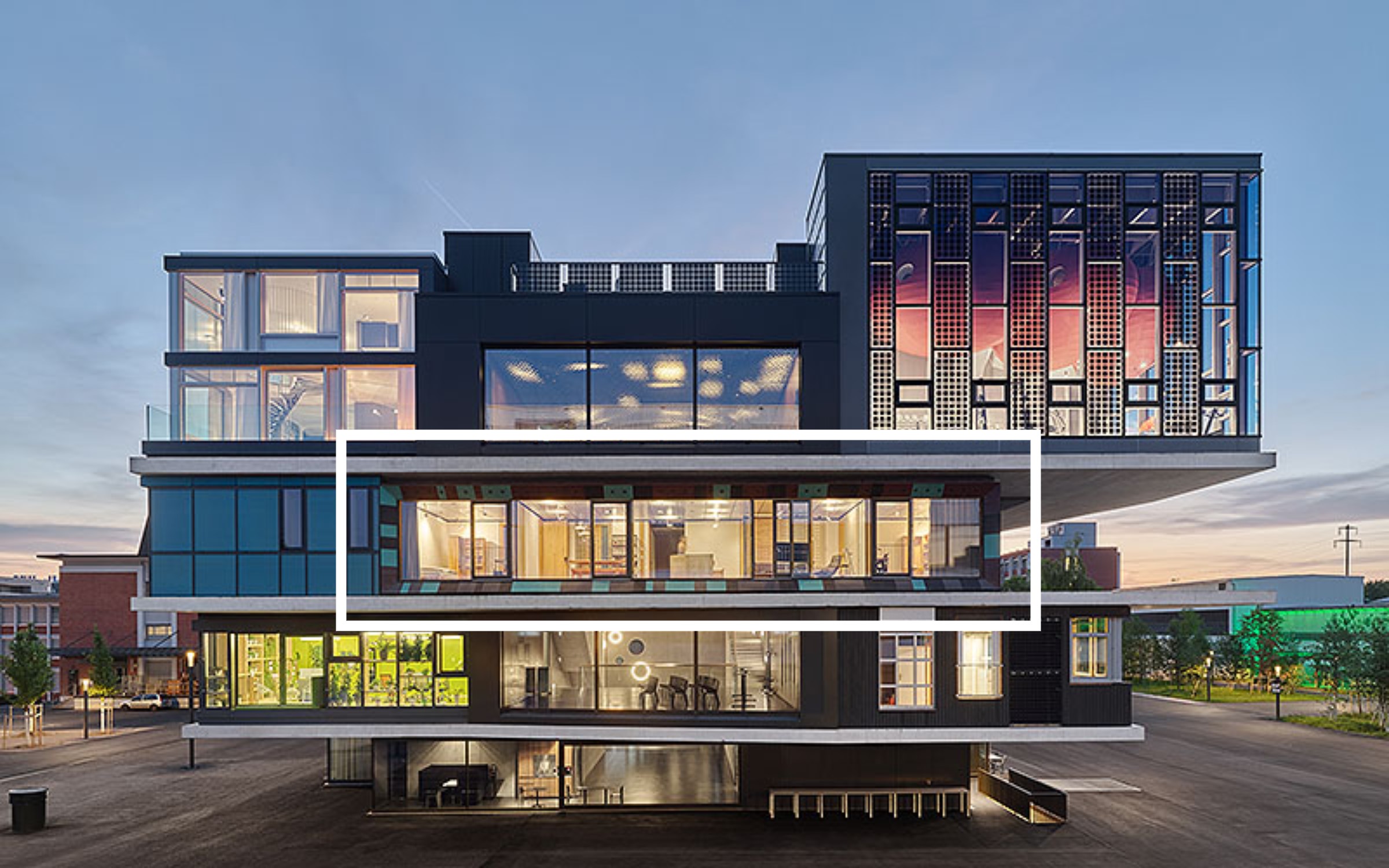}
        \label{Fig:NEST}
    }
    \hfill
    \subfigure[]{ 
        \includegraphics[width=0.225\textwidth]{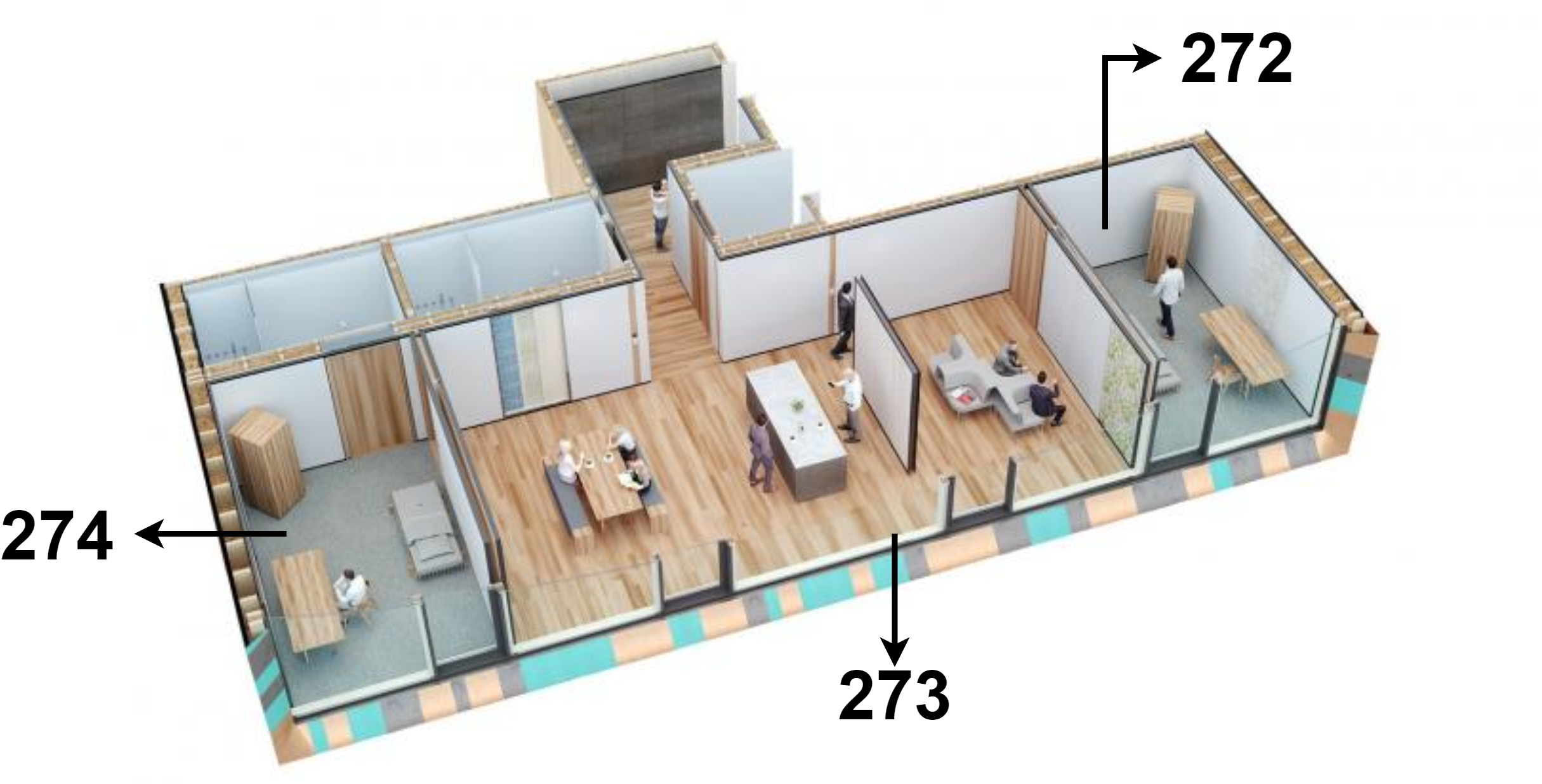}
        \label{Fig:UMARoutline}
    }
    \caption{(a) The NEST building with the UMAR unit marked in white {(© Zooey Braun, Stuttgart)} and (b) Outline of the UMAR apartment unit. The apartment has two bedrooms with a floor area of 17.6~\( \mathrm{m}^2 \) and a large window facing east-southeast. It has a combined heating and cooling system through ceiling panels.}
    \label{Fig:NEST_UMAR}
\end{figure}

The historical data from the \ac{UMAR} unit, a residential space of the Empa NEST Demonstrator in Dubendorf, Switzerland, shown in Fig.~\ref{Fig:NEST_UMAR} are used  \citep{UMAR,Nestweb}. 
We consider the heating of one bedroom, i.e., room 272, to test the proposed control strategy. In this room, we can control the valve to regulate the water flow into the ceiling panels.\\

\subsection{Simulation Setup}
\label{Sec:Settings}

The proposed controller is applied to the building model of the \ac{UMAR} unit, which is trained on historical heating data between December 2021 and March 2022 with 15-minute resolution. The data used are normalized between [0.1, 0.9] using min-max normalization to avoid biased results based on the different scales of the variables.

To evaluate the performance of the proposed \ac{SRL} controller, we define specific conditions for comfort room temperature, electricity pricing, and flexibility provision. The comfort range for room temperature is defined as \([T^{\text{min}}, T^{\text{max}}] = [23.5, 25] ^{\circ} \text{C}\)\footnote{Due to the big windows of UMAR units of NEST facing towards solar irradiation most of the day, the room does not need much heating power, even on colder winter days. Therefore, we chose a higher-than-normal range for the comfort temperature in order to stimulate the controller to take action.} and a dynamic price tariff is used as electricity price. 
The specifications of the flexibility request in the form of \(F = (t_\text{s}, t_\text{e}, \phi_{\text{flexibility}})\) are communicated every day at 8\,am. The start and end times of flexibility windows are randomly selected for the training of the \ac{DRL} agent, with a maximum duration of ten hours, to simulate various flexibility requests. The flexibility coefficient \(\phi_{\text{flexibility}}\) is also randomly selected within the given boundaries \([\phi_{\text{low}}, \phi_{\text{high}}] = [0.7, 1.3]\), corresponding to lower or higher energy consumption compared to the \ac{BAU} scenario.
The weighting factors in the reward function, balancing comfort violations and electricity price, are set to \(\beta = 20\) and \(\delta = 0.8\), while the penalty for not fulfilling flexibility constraints is set to \(\alpha_{1} = 1\) and \(\alpha_{2} = 10\). The weighting factor for tolerance calculation in the \ac{RASF} is chosen as \(w_1 = 0.5\) for this case study.

To evaluate the performance of the proposed \ac{DRL}-based controller with the safety filter, we compare it against a baseline \ac{RB} controller, analyzing the temperature evolution, the electricity cost, and the flexibility provision.
The \ac{RB} controller follows a conventional hysteresis-based strategy. The valve of the heat pump is opened when the room temperature \(T_{\text{room},t}\) falls below the lower comfort boundary \(T^{\text{min}}_t\), and closed when \(T_{\text{room},t}\) exceeds the upper comfort boundary \(T^{\text{max}}_t\). 

As a result, the baseline controller does not optimize heating based on electricity price variations and does not incorporate demand-side flexibility provision. It only reacts to the temperature threshold, leading to heat pump ON/OFF cycles.
The \ac{BAU} scenario, used for flexibility time windows, is computed by projecting future actions based on historical data without assuming flexibility provision or other external disturbances.

\subsection{Numerical Results}
\label{Sec:Results}

This subsection presents the performance of the proposed \ac{DRL}-based control approach with a real-time adaptive safety-filter. We compare four control strategies:

\begin{enumerate}
    \item A baseline \ac{RB} controller based on fixed, manually defined rules for heating control, without consideration of energy cost or flexibility provision.
    \item A \ac{DRL} controller that does not consider flexibility provision. The controller aims to maintain occupant thermal comfort while minimizing electricity costs.
    \item A \ac{DRL} controller that considers flexibility provision but does not include the safety filter.  The system operator communicates flexibility provision messages as introduced in Section \ref{Sec:Flexibility provisions} and Section \ref{Sec:Settings} and the \ac{DRL}-based controller applies its policy to comply with the flexibility constraints. 
    \item A \ac{DRL} controller with the real-time adaptive safety filter to ensure compliance with the flexibility constraints.
\end{enumerate}

\begin{figure*}[hb!]
    \centering
    \includegraphics[width=\textwidth,trim=0 30 0 25, clip]{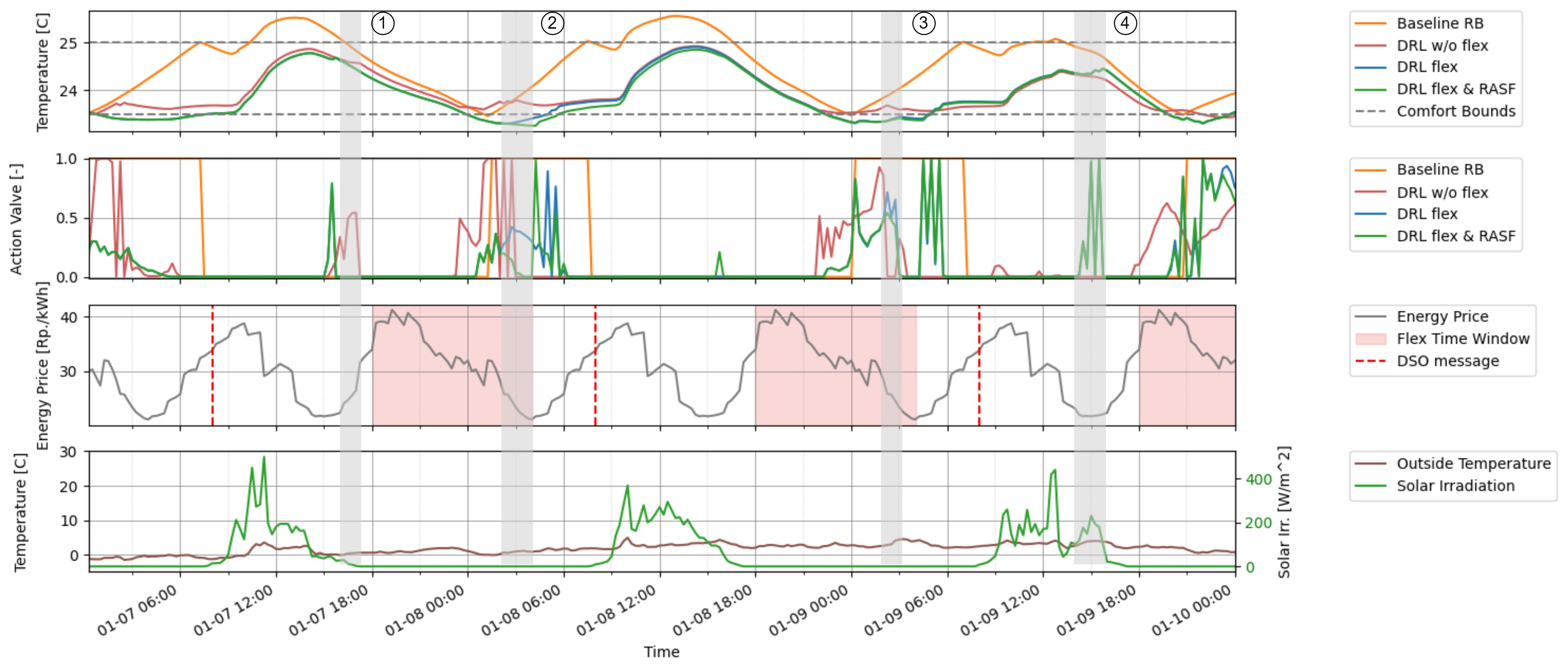}
    \caption{Room temperature evolution, control action and ambient variables under \ac{DRL}-based control with safety filter (green), \ac{DRL}-based control without safety filter (blue), and baseline control (orange) with flexibility time windows (red shaded area) with \(F = [18:00, 04:00, 0.7]\). The grey dashed lines represent the temperature comfort range.}
    \label{fig:ResultsOverview}
\end{figure*}

\begin{figure}[h!]
    \centering
    \includegraphics[width=0.9\linewidth, trim=0 6 0 5, clip]{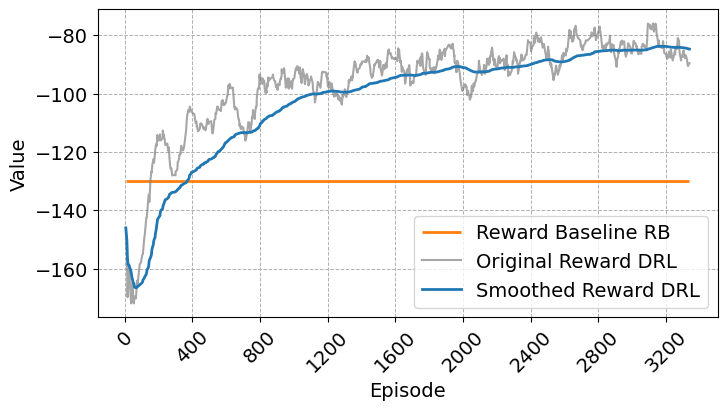}
    \caption{Mean reward of \ac{DRL}-based controller during the training of the agent in Case III compared to the reward of the baseline \ac{RB} controller. For the \ac{DRL} agent by considering flexibility provisions, after about 2'800 episodes, the mean reward curve starts to converge at a reward of approximately -80, where a higher reward value shows better performance in terms of the integrated cost and comfort measure defined in (\ref{eq:rewardfunction}).}
    \label{fig:training}
\end{figure}

\begin{figure}[t]
    \centering
    \includegraphics[width=0.9\linewidth]{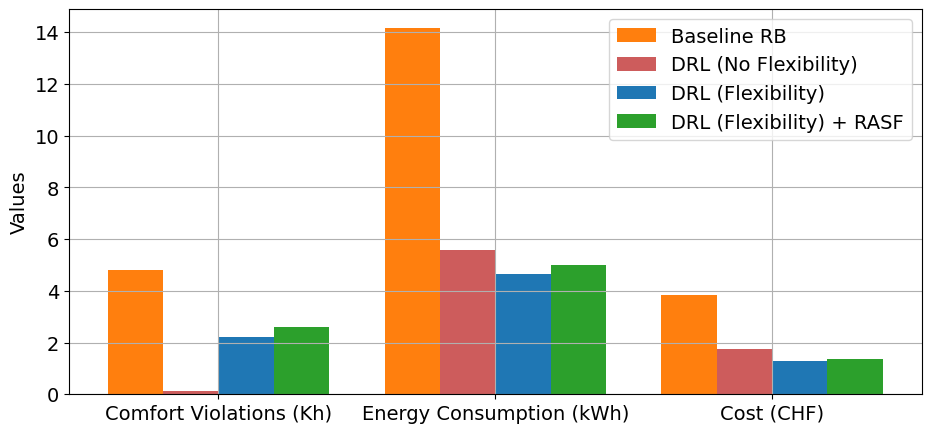}
    \caption{Comparison of Key Performance Indicators over three days.}
    \label{fig:kpi_scenarios}
\end{figure}

Fig.~\ref{fig:ResultsOverview} presents the room temperature evolution, control actions, and ambient conditions for the four different control strategies, highlighting their responses to flexibility requests and varying electricity prices over a three-day period. To complement this, Fig.~\ref{fig:kpi_scenarios} summarizes the key performance indicators across the same simulation period, providing a quantitative comparison of energy consumption, cost, and comfort violations.
Fig.~\ref{fig:training} illustrates the average cumulative reward of the \ac{DRL}-based controller for Case III  during its training over a series of episodes. One episode corresponds to three days. 
It can be seen that the \ac{DRL} controller outperforms the \ac{RB} controller in terms of reward after 400 episodes. 

In Case III, the \ac{DRL} controller is confronted with flexibility constraints. The controller therefore must adjust its operation based on external flexibility disturbances. The flexibility time windows are placed when the agent generally takes action to heat the room (Case II).
We observe that the \ac{DRL}-based controller responds to the flexibility provision request, trying to reduce the energy consumption by 30\% compared to historical data for this time window with \(\phi_{\text{flexibility}} = 0.7\), as shown in Fig.~\ref{fig:ResultsOverview}. As a result, the agent tries to preheat the room before flexibility time windows during low electricity prices in order to comply with flexibility constraints, as seen at \circnum{4}. Compared to the baseline \ac{RB} controller, cost savings are increased heavily, and comfort temperature violations are decreased. However, despite prioritizing the flexibility provision over comfort temperature in the reward function, the controller fails to fulfill all flexibility requests. 

In Case IV, we introduce the \ac{DRL}-based controller combined with the \ac{RASF} introduced in Section~\ref{Sec:Safety Filter}, ensuring strict compliance with the flexibility constraints during flexibility time windows.
In Fig.~\ref{fig:ResultsOverview}, we see that the safety filter ensures fulfillment of the flexibility constraints, for example at \circnum{2} \& \circnum{3}, where the agent without safety filter would not comply with them. Compared to Case III, the comfort temperature violation increases minimally, see \circnum{2}, and energy consumption and cost of electricity are comparable, as shown in Fig.~\ref{fig:kpi_scenarios}. The controller with the safety filter still outperforms the baseline \ac{RB} controller by reducing cost by 64.5\% while improving thermal comfort by 45.5\%.

In Case II, the agent maintains temperatures near the lower comfort bound, as shown in Fig~\ref{fig:ResultsOverview}, reducing fluctuations compared to the \ac{RB} controller. It preheats during low-price periods (see \circnum{1}) to reduce later heating needs, effectively lowering demand during peak prices. As shown in Fig.~\ref{fig:kpi_scenarios}, this results in 54.9\% cost savings and minimal comfort violations. Compared to Cases III and IV, Case II shows higher energy use and cost due to the absence of flexibility constraints.

As seen in Fig.~\ref{fig:kpi_scenarios}, the results show that \ac{DRL}-based controllers significantly reduce energy consumption and cost compared to the baseline \ac{RB} controller. However, enforcing flexibility provision increases comfort temperature violations due to the newly imposed constraints. With the integration of the safety-filter, the performance drops minimally, but ensures compliance with flexibility constraints. Despite its trade-off, Case IV maintains a balance between energy and grid efficiency, ensuring adherence to flexibility requests while keeping costs lower than the baseline. These findings highlight the importance of integrating safety mechanisms in \ac{DRL}-based controllers to balance energy efficiency, comfort, and flexibility provision. Table \ref{tab:kpi_comparison} summarizes the key performance indicators over a one-month simulation for the four cases, confirming the same trends.

\begin{table*}[t]
    \centering
    \caption{Comparison of Key Performance Indicators for Simulation over one Month}
    \label{tab:kpi_comparison}
    \begin{tabular}{p{4cm} p{2cm} p{3cm} p{3cm} p{4cm}}
        \toprule
        \textbf{Metric} & \textbf{Baseline RB} &\textbf{DRL (No Flex. Provision)} & \textbf{DRL (Flex. Provision)} & \textbf{DRL (Flex. Provision) + \ac{RASF}} \\
        \midrule
        Comfort Violations (Kh) & 97.12 & 75.62 & 100.56 & 103.56 \\[4pt]
        Energy Consumption (kWh) & 134.62 & 67.95 & 63.68 & 69.02 \\[4pt]
        Cost (CHF) & 40.75 & 21.41 & 18.07 & 19.42 \\
        \bottomrule
    \end{tabular}
\end{table*}

\subsection{Discussion}

The results demonstrate that the \ac{DRL}-based controller outperforms the \ac{RB} baseline in thermal comfort, energy efficiency, and cost savings — even when subject to flexibility constraints. Its predictive, cost-aware actions help minimize temperature fluctuations and reduce energy consumption and cost by up to 50\% over a one month simulation. While the standalone \ac{DRL} controller performs well, its ability to comply with flexibility constraints is not guaranteed. The integration of the proposed safety filter ensures full compliance with flexibility requests, while maintaining performance close to the unconstrained controller. This demonstrates the filter’s effectiveness in enforcing safety without significantly compromising comfort or efficiency.

Future work should investigate the safety filter’s robustness under varying flexibility patterns, comfort-cost trade-offs, and real-world pricing signals. Additionally, deploying the filter in a distributed multi-agent setting, coordinating across multiple buildings with differing thermal dynamics and occupant preferences, would further enhance its value in grid-responsive applications. Notably, as seen in the simulation over one month, flexibility provision during periods of extreme ambient conditions can lead to increased comfort violations. This suggests that flexibility provision may not be appropriate under such conditions, emphasizing the need for improved coordination and communication strategies between power system operators and buildings.

\section{Conclusion} 
\label{Sec:Conclusion}

In this work, we propose a data-driven safe \ac{DRL}-based method to obtain optimal control policy for the building space heating, considering demand-side flexibility provision. We implement a real-time adaptive safety filter to ensure the fulfillment of flexibility requests from electrical system operators, enabling safe and efficient operation. The \ac{DRL}-based controller, together with the real-time adaptive safety filter, optimizes the control problem to maximize the thermal comfort of the occupants and minimize electricity costs for the heat pump operation.
The proposed \ac{DRL}-based controller outperformed traditional \ac{RB} control, achieving better thermal comfort and lower electricity costs. Combined with the real-time adaptive safety filter, the control architecture successfully fulfills the flexibility constraints, providing the predefined energy flexibility and enhancing the operation of the entire electrical system.
Due to its model-free nature, the approach can be generalized to different buildings, but adaptation requires availability of thermal models and historical data. 

\section*{DECLARATION OF GENERATIVE AI AND AI-ASSISTED TECHNOLOGIES IN THE WRITING PROCESS}
During the preparation of this work, the authors used ChatGPT to check grammar. After using this tool, the authors reviewed and edited the content as needed and take full responsibility for the final version of the publication.
\bibliography{main}             
\end{document}